\documentclass[floatfix,twocolumn,prl]{revtex4}

\usepackage{graphicx}

\begin{document} 

\title{Imaging transport resonances in the quantum Hall effect}

\author{G. A. Steele} 
\author{R. C. Ashoori}
\affiliation{Department of Physics, Massachusetts Institute of Technology, Cambridge, Massachusetts 02139}

\author{L. N. Pfeiffer} 
\author{K. W. West} 
\affiliation{Bell Laboratories, Lucent Technologies, Murray Hill, New Jersey 07974}

\date{June 13, 2005}

\begin{abstract}

We use a scanning capacitance probe to image transport in the quantum Hall system. Applying a DC bias voltage to the tip induces a ring-shaped incompressible strip (IS) in the 2D electron system (2DES) that moves with the tip. At certain tip positions, short-range disorder in the 2DES creates a quantum dot island in the IS. These islands enable resonant tunneling across the IS, enhancing its conductance by more than four orders of magnitude. The images provide a quantitative measure of disorder and suggest resonant tunneling as the primary mechanism for transport across ISs.

\end{abstract}

\maketitle

In the quantum Hall effect (QHE), the Hall resistance of a two-dimensional electronic system (2DES) is quantized with high precision to values that depend only on fundamental physical constants \cite{qhe}. The effect is very robust; it is insensitive to the effective mass, level of disorder, and size of the sample. Arguments by Laughlin \cite{laughlin81} showed that the QHE occurs in any 2D system in which electrons in the bulk of the sample are localized. While the quantization is understood, there is no consensus on a microscopic picture of this localization \cite{huckestein95,network_models}. 

In one picture \cite{network_models}, non-linear screening \cite{efros88} of the 2DES in high magnetic fields plays a central role. Considerations of self-consistent non-linear screening at the edge of the 2DES lead to the prediction of incompressible strips (IS) that isolate regions of different Landau level occupation \cite{chklovskii92}. At quantum Hall plateaus one of the ISs spreads to the central region of the sample to form a percolating network responsible for localization. 

Despite the large body of literature surrounding the QHE, relatively few experiments have studied directly transport across ISs. One experiment measuring transport across a single IS \cite{zhitenev96} displayed sharp transport resonances as a function of a local gate voltage. Chklovskii argued \cite{chlovskii96} that the resonances arose from resonant tunneling through Coulomb blockaded compressible islands within the IS. A closely related phenomenon arises in narrow Hall bars. Here, a single IS separates the two edges of the sample. In contrast to the smooth transitions between Hall plateaus seen in large samples, narrow Hall bars show sharp reproducible fluctuations \cite{cobden99,fluctuations2}. In analogy with Chklovskii's arguments, Cobden {\em et al.\ }suggested that these fluctuations could arise from resonant tunneling through Coulomb blockaded islands. However, a well defined oscillation period typical of Coulomb blockade was not seen, and other models have been presented for the observed fluctuations \cite{machida01_short}. Finally, measurements of local chemical potential fluctuations \cite{ilani04_short} now demonstrate the existence of such disorder-induced Coulomb blockaded islands in quantum Hall systems but do not address their influence on transport.

In this letter, we present results from a new method that images charge transport across ISs on a microscopic scale. We find that resonant tunneling through Coulomb blockaded islands located inside the IS is the chief means by which electrons traverse ISs. The islands create more than a 10,000 fold enhancement of the transport current compared to the situation where no island exists in the strip. Finally, the data reveal the shapes of ISs formed by a tip perturbation and the 2DES disorder and provide a detailed picture of disorder in high mobility samples.

\begin{figure} 
\includegraphics[height=2.0in]{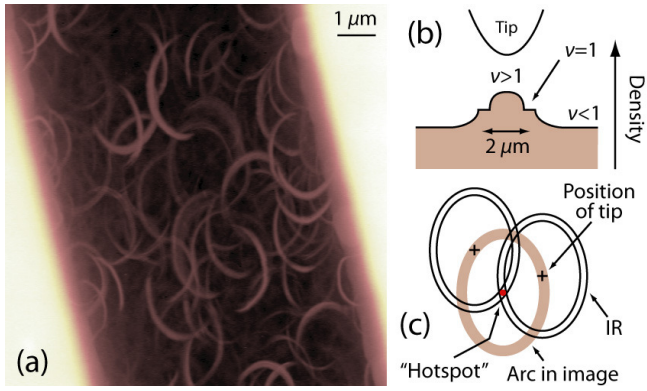}
\caption{\label{fig1} (color online)
(a) 10x10 $\mu$m in-phase charging image taken at $\nu_{bulk}=0.89$ (B=7.0T) with a bias of $+1.5$V applied to the tip. (b) Bubble induced by DC tip bias, surrounded by an incompressible ring (IR). Scale bar indicates the size of the IR for the image shown. (c) Intersection of the IR with a ``hotspot'' of fixed position in the 2DES creates filamentary arcs in the image that reproduce the shape of the IR inverted through the hotspot.}
\end{figure}

We use a scanning charge accumulation (SCA) microscope \cite{tessmer98_short} to probe the resistance of a tip-induced mobile IS. By applying a DC voltage to our tip, we can create the situation shown in Figure \ref{fig1}(b). When the bulk filling factor is less than integer, a ``bubble'' of electrons is induced in the next Landau level underneath the tip. The bubble is separated from the surrounding 2DES by a ring shaped IS that we refer to as an ``incompressible ring'' (IR). By capacitively detecting the charging of the bubble in response to an AC excitation applied the sample, we directly measure the resistance of the IR. Moving the tip, the bubble and ring translate together, producing an image that is a map of the resistance of IRs formed at different locations in the sample.

Experiments are performed in a $^3$He cryostat at a temperature of 300 mK. Our sample is an AlGaAs/GaAs heterostructure with a 10 nm thick metal finger gate \cite{glicofridis02} patterned on the surface. A 2DES with a density of $1.5 \times 10^{11}$ cm$^{-2}$ and a mobility of $4.5 \times 10^6$ cm$^2$/Vs is located 100 nm below the surface. A 15 mV AC excitation at 200 kHz is applied to both the 2DES and the gate while the tip is scanned at a fixed height of 40 nm above the surface. As the bubble is connected only capacitively to ground, the AC voltage appearing across the IR is reduced by a capacitive lever-arm. Our simulations show a factor of 10 reduction \cite{simulation}. Experimentally, reducing the excitation does not substantially alter the images. The DC voltages applied to the tip are referenced to the voltage that nulls the electric field due to the work function difference between the tip and the 2DES. In order to increase the capacitive signal from the charging of the bubble, we image using a tip that has a 2 $\mu$m radius of curvature. This does not decrease our effective spatial resolution, which is set by the width of the IR.

Figure \ref{fig1}(a) displays an image of the in-phase component of the charging signal, taken with the 2DES at filling factor $\nu_{bulk}=0.89$ and a positive bias voltage applied to the tip. The gates appear as regions of high capacitance in the top right and bottom left areas of the image. In the region of the image where the tip is over the 2DES, a bubble forms from a local accumulation of electrons in the $\nu>1$ state. The image is generally dark due to the large resistance of the surrounding IR, with the exception of sharp, bright elliptical arcs. The patterns repeat at uncorrelated positions in the image. In the absence of a magnetic field, the images are completely featureless. The observed arcs appear only at magnetic fields near integer filling factors and upon applying a bias voltage of an appropriate polarity for the formation of a bubble.

\begin{figure}
\includegraphics[height=2.0in]{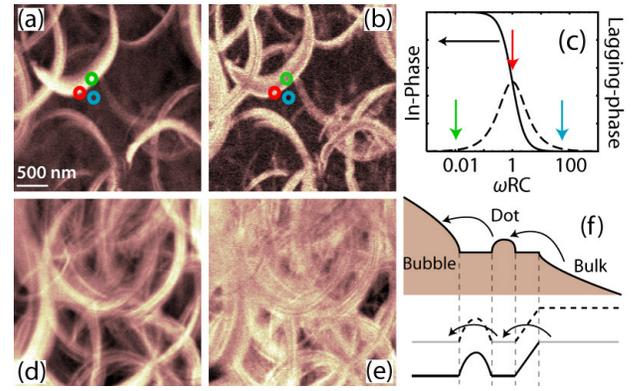}
\caption{\label{fig2} (color online)
3x3 $\mu$m charging images. (a),(b) In-phase and lagging phase electron-bubble images, $\nu_{bulk}=0.83$ (7.5T), $V_{tip}=+1.5$V. Circles drawn in the images show tip positions where the IR resistance varies to result in the in-phase and lagging-phase signals shown by matching arrows in the RC model in (c). (d),(e) In-phase and lagging-phase hole-bubble images for $\nu_{bulk}=1.1$ (5.5T), $V_{tip}=-1.5$V. (f) Density profile (top) and Landau level energy profile (bottom) of a short length-scale density fluctuation creating a quantum dot allowing resonant tunneling across the IR. Dotted, gray and solid lines (bottom) depict empty, partially filled and filled Landau levels respectively. Coulomb blockade of the dot leads to sub-filament fringes visible in the images.}
\end{figure}

We interpret the images in the following model. The bright arcs correspond to the locus of tip positions at which the IR intersects a ``hotspot'' in the 2DES, as illustrated in figure \ref{fig1}(c). The hotspot originates from a short length-scale density fluctuation in the 2DES, which creates a small quantum dot (or anti-dot) island embedded in the IR, as illustrated in figure \ref{fig2}(f). Electrons can then resonantly tunnel across the IR through the dot, causing the IR resistance to drop dramatically. With this resonant tunneling, the bubble charges on time scales comparable or faster than the inverse of the measurement frequency. The arcs in the image directly reflect the shape of the IR created by the tip, and the density fluctuations (hotspots) responsible for creating the dots are located at the centers of the ellipses suggested by the arcs. The size of the IR depends on the magnetic field and tip bias voltage, and can be measured from the size of arcs in the image. The elliptical shape is a result of asymmetry in the tip etching procedure.

Our data fit a simple RC model in which the capacitance of the bubble to its surroundings charges through the resistance of the IR. Figure \ref{fig2}(c) depicts the expected behavior in this model of the in-phase and lagging-phase signals as a function of the RC charging time of the bubble. Figures \ref{fig2}(a) and (b) show in-phase and lagging phase images taken at a high spatial resolution (3x3 $\mu$m area).  At the location marked by the green circles, the IR has low resistance and the bubble nearly fully charges during the cycle of the excitation. The blue circles correspond to at least 10,000 times higher IR resistance and the bubble does not charge at all. 

Many filamentary arcs in the images also show narrow sub-filament oscillations, which we attribute to Coulomb blockade. The dot combined with the bubble acts as a single-electron transistor \cite{kastner92} (SET), with the bulk 2DES as the source contact and the bubble as the drain contact. The SET island is embedded in the IR, and is gated by electric fields from the tip, leading to discrete changes in its occupancy as the tip is moved, or as the DC bias is changed with the tip at a fixed position.
 
We also probe ISs by moving to a field with $\nu_{bulk}>1$ and reversing the bias voltage on the tip to produce an accumulation of holes in the lower filled Landau level (a ``hole bubble''). This allows us to image the resistance of an IR formed at a lower magnetic field. As shown figures \ref{fig2}(d) and (e), the hole bubble IR has a lower resistance and the images display many more filaments, indicating a reduced tunnel barrier formed from the exchange enhanced spin gap at these magnetic fields. We also observe filaments at higher filling fractions. IRs formed by orbital energy gaps at even filling factors $\nu=2$ and $\nu=4$ display a much larger resistance than those at exchange enhanced spin gaps at $\nu=1$ and $\nu=3$.  

It is important to emphasize that we believe that the hotspots are not simply ``defects'' in the 2DES. Different filaments in the images show resonant conductance enhancements that vary over orders of magnitude suggesting they are not associated with an impurity of fixed strength. The hotspots instead result from short length-scale fluctuations that are present everywhere in high mobility heterostructures with remote ionized donors. At zero magnetic field, such short length-scale fluctuations lead only to small angle scattering and smooth branching electron flow \cite{topinka01_short} because the amplitude of the fluctuations $\Delta U$ is small compared to the Fermi energy $E_F$. In contrast, our work shows that at high magnetic fields these weak short length-scale fluctuations have a drastically different effect, creating Coulomb blockaded islands that enable resonant transport of charge across ISs.

\begin{figure}
\includegraphics[width=3.1in]{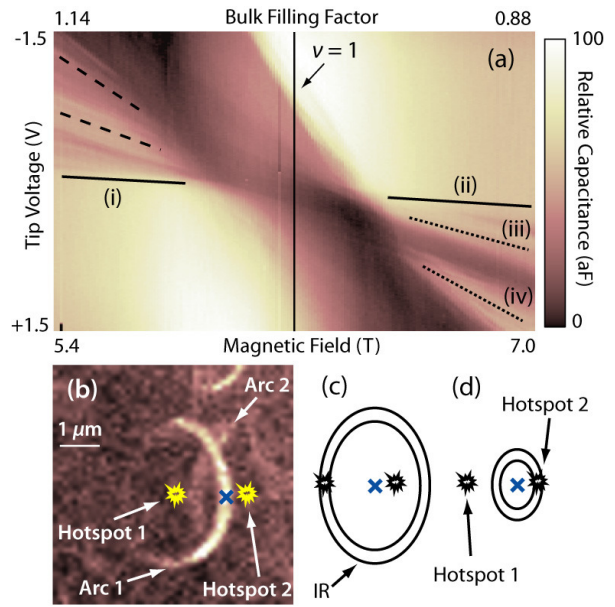}
\caption{\label{fig3} (color online)
(a) Color scale plot of the in-phase charging signal as a function of tip bias voltage and magnetic field with the tip at a fixed position. Lines (i) and (ii) display the threshold for forming a hole and electron bubble respectively. Dashed (dotted) lines indicate biases and fields where the IR intersects a hotspot, allowing resonant tunneling into a hole (electron) bubble under the tip. (b) 6x6 $\mu$m in-phase electron bubble image at 7.0T and +1.5V. The blue marker shows the position of the tip during the measurement in (a). Two arcs are identified, and the positions of the hotspots responsible for each arc are indicated. (c) The IR intersects hotspot 1 at biases and fields given by line (iv) in (a). (d) At lower biases the IR is smaller, and intersects hotspot 2 at biases and fields given by line (iii).}
\end{figure}

Varying the tip bias voltage causes the arcs in the images to grow or shrink in size. For an electron bubble, making the bias voltage more positive increases the number of electrons in the $\nu>1$ region under the tip. This causes the IR to grow in size leading to larger observed arcs. Similarly, increasing the magnetic field for an electron bubble decreases the occupation of the bubble, and the IR and imaged arcs shrink. To characterize this behavior, we study the bias voltage and magnetic field dependence of the charging signal with the tip at a fixed position, shown in figure \ref{fig3}. The drop in signal appearing in the lower right (upper left) regions of figure \ref{fig3} arises from the formation of an electron (hole) bubble IR. Making the bias voltage more positive (negative), the IR grows and peaks appear in the charging signal as the IR intersects different hotspots. Changing the magnetic field, the hotspot peaks follow sloped lines in the bias-field plot that converge on null voltage when $\nu_{bulk}=1$. Near the center of the $\nu_{bulk}=1$ Hall plateau, the bulk conductivity of the sample vanishes, and there is no signal because charge cannot penetrate the sample.

The lagging-phase data shown in figures \ref{fig2}(b) and (e) demonstrate that the positions at which the bubble does not charge correspond to RC charging times that are far in excess of the period of the AC measurement. We roughly estimate the self capacitance of the bubble from its size: a 2 $\mu$m diameter bubble in GaAs gives a self capacitance of 1 fF. Estimating the product $\omega$RC to be at least 100 from the data gives a value of at least 100 G$\Omega$ for the resistance of a pristine IR.

The charge fluctuations that produce the quantum dot islands in the IR involve length-scales smaller than the width of the IR. To estimate the IR width, we have completed a series of electrostatic simulations that account for the geometry of the tip as well as the nonlinear screening of the 2DES \cite{simulation}. From these, we obtain an IR width of 200--300 nm in the absence of disorder. Density fluctuations on length-scales larger than the IR width appear as an additional local density gradient superimposed on the density gradient from the tip. The width of an IS is inversely proportional to the density gradient creating it \cite{chklovskii92}. Consequently, the local IR width will vary depending on the relative orientation of the gradients from the tip and the disorder.

\begin{figure}
\includegraphics[width=3.05in]{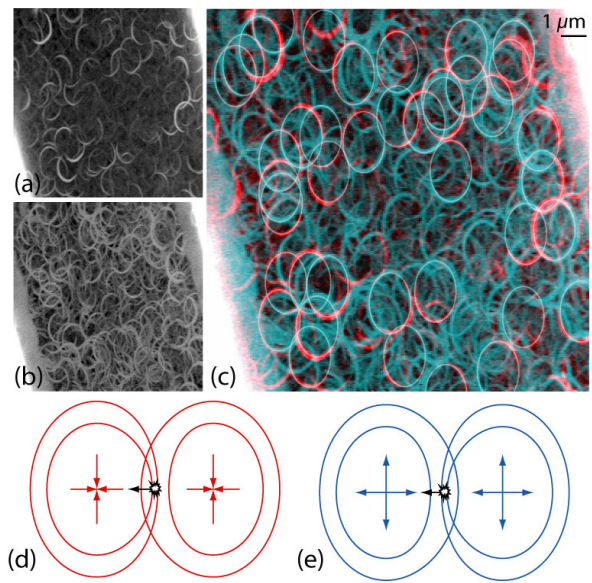}
\caption{\label{fig4} (color online)
(a) 15x15 $\mu$m in-phase electron-bubble image, taken at $\nu_{bulk}=0.83$ (7.5T), $V_{tip}=+1.75$V. (b) 15x15 $\mu$m in-phase hole-bubble image at the same location at $\nu_{bulk}=1.26$ (5.0T). A tip bias of $-1.75$V is chosen to produce an IR of the same diameter as in (a). (c) Composite image constructed from (a) and (b) with hole-bubble data in blue and electron-bubble data in red. Semi-transparent repeating ellipses are overlaid as a guide to the eye. (d),(e) Schematics illustrating local IR width modulation from disorder in (d) electron and (e) hole bubbles. The black marker and arrow indicate the location of a hotspot and direction of a fixed local density gradient. Colored arrows indicate the direction of the density gradient from the tip.}
\end{figure}

Analysis of figure \ref{fig4} explains our observation of only partial ellipses in the context of such local IR width modulations. We compare electron and hole bubble images taken at the same location. In switching from an electron to a hole bubble, the local density gradient from the tip inverts. To obtain the same local IR width at the location of the hotspot, the tip must be moved to the opposite side of the hotspot, as shown in figures \ref{fig4}(d) and (e). Figure \ref{fig4}(c) shows a composite image formed from an electron bubble image and a hole bubble image taken at the same location. In places where a partial ellipse appears in the electron bubble image, a partial ellipse in the hole bubble image appears on the opposite side of the hotspot. This remarkable symmetry between the two images demonstrates that the partial ellipses arise from the influence of longer length-scale disorder.

Including the effects of disorder on all length-scales, we would expect that rather than a smooth ellipse as shown in figure \ref{fig1}(c), the IR formed by the density gradient from the tip would trace out a path that meanders between hills and valleys in the disorder potential, and that would change shape at different locations in the sample. In contrast, images display only smooth arcs that fall on the same repeating ellipse. The disorder in this high mobility sample appears to have the character of being relatively uniform over long length-scales with sharp fluctuations on small length-scales. These predictions can be made quantitative through comparison with simulations that include remote ionized donors \cite{simulation}.

In summary, we have observed strong resonances in transport across ISs. Using local imaging, we have determined that these resonances arise from the intersection of the strip with fixed positions in the 2DES and are associated with the formation of a small quantum dot island embedded in the incompressible region. These islands are created by disorder, and imaging transport through them allows us to infer the nature of the disorder in the 2DES. On small length-scales ($\sim$100 nm), significant disorder exists, creating the islands in the IR. Fluctuations on an intermediate scale are responsible for our observation of only partial rings in the images. On larger length-scales ($\sim$1 um), the amplitude of the disorder fluctuations is relatively small, as evidenced by the uniformity of the size and shape of the rings in the images. 

Finally, the quantum dot islands created by the disorder have a dramatic effect on charge transport across ISs. They provide at least 10,000 times the conductance of tunneling directly across our IR, or equivalently, the same conductance of an ideal IS that is more than 3 cm long. Their presence opens a conduction channel that dominates all transport across narrow incompressible regions. This observation is in agreement with ideas presented by Cobden {\em et al.\ }\cite{cobden99} for the origin of fluctuations observed in narrow devices. Moreover, the impressive magnitude of the conductance enhancement suggests that this resonant tunneling may act as the fundamental mechanism for transporting charge through the network of ISs in large samples, mediating hops between larger compressible islands. To our knowledge, such a resonant tunneling mechanism, distinct from variable range hopping or thermal activation \cite{rxxminima}, has not been considered in theories describing transport around $\rho_{xx}$ minima.

We would like to thank N. L. Spasojevic for his work in developing the electrostatic simulations. This work was supported by the Office of Naval Research, and by the National Science Foundation through the NSEC and MRSEC programs.


\begin{thebibliography}{17}
\expandafter\ifx\csname natexlab\endcsname\relax\def\natexlab#1{#1}\fi
\expandafter\ifx\csname bibnamefont\endcsname\relax
  \def\bibnamefont#1{#1}\fi
\expandafter\ifx\csname bibfnamefont\endcsname\relax
  \def\bibfnamefont#1{#1}\fi
\expandafter\ifx\csname citenamefont\endcsname\relax
  \def\citenamefont#1{#1}\fi
\expandafter\ifx\csname url\endcsname\relax
  \def\url#1{\texttt{#1}}\fi
\expandafter\ifx\csname urlprefix\endcsname\relax\def\urlprefix{URL }\fi
\providecommand{\bibinfo}[2]{#2}
\providecommand{\eprint}[2][]{\url{#2}}

\bibitem[{\citenamefont{Prange and Girvin}(1990)}]{qhe}
\bibinfo{author}{\bibfnamefont{R.~R.} \bibnamefont{Prange}} \bibnamefont{and}
  \bibinfo{author}{\bibfnamefont{S.~M.} \bibnamefont{Girvin}},
  \emph{\bibinfo{title}{The Quantum Hall Effect}}
  (\bibinfo{publisher}{Springer}, \bibinfo{address}{New York},
  \bibinfo{year}{1990}).

\bibitem[{\citenamefont{Laughlin}(1981)}]{laughlin81}
\bibinfo{author}{\bibfnamefont{R.~B.} \bibnamefont{Laughlin}},
  \bibinfo{journal}{Phys. Rev. B} \textbf{\bibinfo{volume}{23}},
  \bibinfo{pages}{R5632} (\bibinfo{year}{1981}).

\bibitem[{\citenamefont{Huckestein}(1995)}]{huckestein95}
\bibinfo{author}{\bibfnamefont{B.}~\bibnamefont{Huckestein}},
  \bibinfo{journal}{Rev. Mod. Phys.} \textbf{\bibinfo{volume}{67}},
  \bibinfo{pages}{357} (\bibinfo{year}{1995}).

\bibitem[{net()}]{network_models}
N.\ R.\ Cooper and J.\ T.\ Chalker, Phys.\ Rev.\ B {\bf 48},
  4530 (1993); D.\ B.\ Chklovskii and P.\ A.\ Lee, Phys.\ Rev.\ B {\bf 48},
  18060 (1993).

\bibitem[{\citenamefont{Efros}(1988)}]{efros88}
\bibinfo{author}{\bibfnamefont{A.~L.} \bibnamefont{Efros}},
  \bibinfo{journal}{Solid State Commun.} \textbf{\bibinfo{volume}{67}},
  \bibinfo{pages}{1281} (\bibinfo{year}{1988}).

\bibitem[{\citenamefont{Chklovskii et~al.}(1992)\citenamefont{Chklovskii,
  Shklovskii, and Glazman}}]{chklovskii92}
\bibinfo{author}{\bibfnamefont{D.~B.} \bibnamefont{Chklovskii}},
  \bibinfo{author}{\bibfnamefont{B.~I.} \bibnamefont{Shklovskii}},
  \bibnamefont{and} \bibinfo{author}{\bibfnamefont{L.~I.}
  \bibnamefont{Glazman}}, \bibinfo{journal}{Phys. Rev. B}
  \textbf{\bibinfo{volume}{46}}, \bibinfo{pages}{15606} (\bibinfo{year}{1992}).

\bibitem[{\citenamefont{Zhitenev et~al.}(1996)\citenamefont{Zhitenev, Brodsky,
  Ashoori, and Melloch}}]{zhitenev96}
\bibinfo{author}{\bibfnamefont{N.~B.} \bibnamefont{Zhitenev}},
  \bibinfo{author}{\bibfnamefont{M.}~\bibnamefont{Brodsky}},
  \bibinfo{author}{\bibfnamefont{R.~C.} \bibnamefont{Ashoori}},
  \bibnamefont{and} \bibinfo{author}{\bibfnamefont{M.~R.}
  \bibnamefont{Melloch}}, \bibinfo{journal}{Phys. Rev. Lett.}
  \textbf{\bibinfo{volume}{77}}, \bibinfo{pages}{1833} (\bibinfo{year}{1996}).

\bibitem[{\citenamefont{Chklovskii}()}]{chlovskii96}
\bibinfo{author}{\bibfnamefont{D.~B.} \bibnamefont{Chklovskii}},
  \eprint{cond-mat/9609023}.

\bibitem[{\citenamefont{Cobden et~al.}(1999)\citenamefont{Cobden, Barnes, and
  Ford}}]{cobden99}
\bibinfo{author}{\bibfnamefont{D.~H.} \bibnamefont{Cobden}},
  \bibinfo{author}{\bibfnamefont{C.~H.~W.} \bibnamefont{Barnes}},
  \bibnamefont{and} \bibinfo{author}{\bibfnamefont{C.~J.~B.}
  \bibnamefont{Ford}}, \bibinfo{journal}{Phys. Rev. Lett.}
  \textbf{\bibinfo{volume}{82}}, \bibinfo{pages}{4695} (\bibinfo{year}{1999}).

\bibitem[{flu()}]{fluctuations2}
See also E.\ Peled {\em et al.}, Phys.\ Rev.\ Lett.\ {\bf 91},
  236802 (2003) and references therein.

\bibitem[{mac()}]{machida01_short}
T.\ Machida {\em et al.}, Phys.\ Rev.\ B {\bf 63} 045318
  (2001).

\bibitem[{ila()}]{ilani04_short}
S.\ Ilani {\em et al.}, Nature {\bf 427} 328 (2004).

\bibitem[{tes()}]{tessmer98_short}
S.\ H.\ Tessmer {\em et al.}, Nature {\bf 392} 51 (1998).

\bibitem[{gli()}]{glicofridis02}
P.\ I.\ Glicofridis, G.\ Finkelstein, R.\ C.\ Ashoori, M.\ Shayegan, Phys.\ Rev.\ B {\bf 65} 121312(R) (2002)

\bibitem[{sim()}]{simulation}
G.\ A.\ Steele {\em et al.}, in preparation.

\bibitem[{\citenamefont{Kastner}(1992)}]{kastner92}
\bibinfo{author}{\bibfnamefont{M.~A.} \bibnamefont{Kastner}},
  \bibinfo{journal}{Rev. Mod. Phys.} \textbf{\bibinfo{volume}{64}},
  \bibinfo{pages}{849} (\bibinfo{year}{1992}).

\bibitem[{top()}]{topinka01_short}
M.\ A.\ Topinka {\em et al.}, Nature {\bf 410} 183 (2001).

\bibitem[{rxx()}]{rxxminima}
See {\em e.g.} D.\ G.\ Polyakov and B.\ I.\ Shklovskii, Phys.\
  Rev.\ Lett.\ {\bf 70} 3796 (1993); {\bf 73} 1150 (1994).

\end{thebibliography}
\end{document}